# New method for obtaining complex roots in the semiclassical coherent-state propagator formula.

*A L Xavier, Jr.*[&]
*Center for Research and Technology,*
*Salesian University Center, Av Almeida Garret 267*
*CEP 13087 290, Campinas, SP Brazil.*

## Abstract

*A semiclassical formula for the coherent-state propagator requires the determination of specific classical paths inhabiting a complex phase-space through a Hamiltonian flux. Such trajectories are constrained to special boundary conditions which render their determination difficult by common methods. In this paper we present a new method based on Runge-Kutta integrator for a quick, easy and accurate determination of these trajectories. Using nonlinear one dimensional systems we show that the semiclassical formula is highly accurate as compared to its exact counterpart . Further we clarify how the phase of the semiclassical approximation is correctly retrieved under time evolution.*

[&] e-mail: *xavnet2@yahoo.com*

## I Introduction

Semiclassical methods have a long history in the study of several quantum systems. These methods (also termed quasi-classical) are able to describe the behavior of quantum states in a certain representation as expansions in terms of classical behavior. There is a characteristic limit for this expansion in which the associated action of the system is sufficiently larger than $\hbar$. However, since the characterization of quantum states strongly depends on representation, the semiclassical limit (or the semiclassical approximation to the state) is implemented by as numerous methods as existing state representations.

Interesting questions arise when a representation making use of phase-space is introduced [1][2]. As we know, such representation can not be unique once the uncertainty principle precludes the complete description of states by canonical conjugate variables. However, phase-space representations can indeed provide useful information on the system dynamics mainly if we take their semiclassical approximation (or if we consider their outcome in the limit $\hbar \to 0$).

The concept of coherent-state was created in 1926 by E. Schrödinger [3] while the systematic study of these states was carried out much later in quantum optics [4], due to the dynamical equivalence between the harmonic oscillator states and the quantized oscillation modes of the electromagnetic field. In terms of the oscillator states, $|n\rangle$, a coherent-state is written as

$$|z\rangle = \exp\left(-\frac{1}{2}|z|^2\right) \sum_{n=0}^{\infty} \frac{z^n}{\sqrt{n!}} |n\rangle, \qquad (1)$$

where $z$ is the coherent state label related to the eigenvalue

$$z = \frac{1}{\sqrt{2}}\left(\frac{q}{b} + i\frac{p}{c}\right), \qquad (2)$$

of the creation and desctruction state operators [5]. In (2), $q$ and $p$ are the average position and momentum of the state (1) and $b$ and $c$ are the respective uncertainties such that

$$bc = \hbar. \qquad (3)$$

The dynamics of quantum systems (and a preparation for the semiclassical study that follows) is well represented by the so called coherent-state propagator (CSP) which is just the coherent-state matrix elements of the time evolution operator

$$K(z'',z',t''-t') = \langle z''|\exp\left(-i\frac{\hat{H}(t''-t')}{\hbar}\right)|z'\rangle, \quad (4)$$

where $\hat{H}$ is the system Hamiltonian operators and $t''-t'$ is the time interval between the initial and final states in consideration. Coherent-states are but a class of the so-called continuous representations [6]. The elements described in (4) contain the whole quantum dynamics associated to the system. For instance, an initial state can be easily evolved in time by using (4). On the other hand, the Fourier transform of the diagonal elements of (4) gives an analytical function whose poles are the energy levels of the system.

In the context of the formalism represented by the CSP, an interesting program is the search for a semiclassical approximation to (4) [6] in the light of similar works [7] which could provide state transition amplitudes built in terms of an underlying classical skeleton. Such semiclassical approximation formula exists [8][9][10] and, as we will show is able to provide accurate values for the CSP even well deep in the purely quantum regime, that is, for large values of $\hbar$.

The organization of the paper is as follows: in section II we give the main elements for the semiclassical coherent-state propagator (SCSP) formula emphasizing the role of the complex roots. In section III we describe a new method of determination of complex trajectories by a simple Runge-Kutta integrator. In section IV we discuss the behavior of the SCSP phase factor under time evolution and give some examples of SCSP calculation for one dimensional systems (harmonic and quartic wells). Finally, in section V we discuss the main results and a future continuation of the work.

## II Semiclassical formulas for the CSP.

In a previous paper [10] we have presented the main relations that compose the semiclassical approximation to the CSP. The aim of that work was to provide a numerical program to the determination of the complex roots in terms of which the SCSP, $\tilde{K}(z'',z',T)$, could be calculated along the time interval $T = t''-t'$. The core element of the SCSP is the determination of complex trajectories labeled by the variables $u$ and $v$ which are functions of the complex conjugate pair of numbers $z$ and $z^*$. The SCSP approximation is given by

$$\tilde{K}(z'',z',T) = \exp\left(-\frac{1}{2}|z''|^2 - \frac{1}{2}|z'|^2\right) \sum_k \sqrt{\frac{1}{\Delta_k}}$$
$$\exp\left(\frac{i}{\hbar}S_k(v'',u',T) + i\mathbf{s}_k\right) \quad (5)$$

Relation (5) was obtained under the limit $\hbar \to 0$ but is in fact valid if $\hbar$ is small in comparison with other quantities of the system (for instance the integral of $pdq$ along a certain path in phase-space), and $v''$ and $u'$ stands for $z''^*$ and $z'$ respectively. Also $\mathbf{s}_k$ is a phase-factor gained after the time evolution which we will discuss later (section IV). The sum $k$ is made over terms that are functions of complex paths bounding the initial and final states $|z'\rangle$ and $|z''\rangle$. These complex trajectories are best described in terms of complex variables $u$ and $v$ evolving according to Hamilton equations

$$i\hbar\dot{u} = \frac{\partial \tilde{H}}{\partial v}, \qquad i\hbar\dot{v} = -\frac{\partial \tilde{H}}{\partial u} \quad (6)$$

which obey the boundary conditions $u' = z'$ and $v'' = z''^*$. The function $\tilde{H}$ in Eqs (6) is the "smoothed" Hamiltonian

$$\tilde{H}(q,p) = \langle z|\hat{H}(q,p)|z\rangle. \quad (7)$$

To complete the description of Eq. (5), the function $S_k(v'',u',T)$ is the generalized action of the k-th contributing complex trajectory and is given by

$$S(v'',u',T) = \int_0^T \left[\frac{i\hbar}{2}(v\dot{u} - u\dot{v}) - \tilde{H}\right]dt + $$
$$-\frac{i\hbar}{2}(v''u'' + v'u') \quad (8)$$

Also, in Eq. (5) we have

$$\Delta_k = -i\hbar\left(\left|\frac{\partial^2 S_k}{\partial u'\partial v''}\right|\right)^{-1} \exp\left(-\frac{i}{\hbar}\int_0^T \frac{\partial^2 \tilde{H}}{\partial u\partial v}\bigg|_k dt\right) \quad (9)$$

$S_k(v'',u',T)$, $\widetilde{H}$ and $\Delta_k$ are complex functions. The dynamics generated by Eqs. (6) inhabits a complex phase space where both position and momentum have real and imaginary parts. This is a necessary requirement of the semiclassical approximation $\widetilde{K}(z'',z',T)$ since, from the point of view of classical "real" mechanics, it is not always possible to find a classical trajectory linking $z'$ to $z''$ in time $T$. There are simply too many boundary conditions to be satisfied. By extending the dynamics to a complex algebra, it is possible to accommodate the additional conditions. Therefore, Eq. (5) gives the probability amplitude of transition between states labeled by $z'$ and $z''$ that are classically connected by complex trajectories.

The SCSP was successfully calculated for a variety of one dimensional systems [11]. In particular, its was possible to propose a new tunneling time in scattering problems [12] where the dwelling time was simply taken as the time interval of the complex trajectory within the potential barrier.

## III New numerical method for calculating the complex roots of the SCSP.

Previously [11] we have proposed a method based on the monodromy matrix [13] for the determination of complex trajectories of the SCSP. Given an initial tentative trajectory, this method iterates the entire initial guess for a number of times according to the linearized dynamics as in the usual Newton´s method. Convergence is attained when the final corrected trajectory satisfies Eq. (6) within a given accuracy range. The greatest advantage of this method was the determination of the SCSP amplitude factor, Eq. (9), by the monodromy matrix elements. However, the monodromy method is not straightforwardly implemented, besides suffering from strong lack of convergence in some situations when the initial guess orbit is not sufficiently close to the answer.

Hamilton equations, Eqs. (6), suggests that a next point predictor method could be used [14], the only difficulty being how to fulfill the unusual boundary conditions $u' = z'$ and $v'' = z''^*$. Here we present a new method which implements this idea. In what follows we work out the main relations for one dimensional systems only.

Let us denote the complex phase-space variables by

$$q = x_1 + ip_2,$$
$$p = p_1 + ix_2, \quad (10)$$

where $x_1$, $p_2$, $p_1$ and $x_2$ are real numbers. If we restrict ourselves to analytical Hamiltonians [11], we can show that Eqs. (6) become

$$\dot{x}_1 = \frac{\partial \widetilde{H}_1}{\partial p_1}, \qquad \dot{x}_2 = \frac{\partial \widetilde{H}_1}{\partial p_2},$$
$$\dot{p}_1 = -\frac{\partial \widetilde{H}_1}{\partial x_1}, \qquad \dot{p}_2 = -\frac{\partial \widetilde{H}_1}{\partial x_2}. \quad (11)$$

We see that the one dimension problem (two degrees of freedom in classical phase-space) was extended to a two dimensional one (four degrees of freedom in complex phase-space). In Eqs. (11) we have

$$\widetilde{H}_1 = \Re[\widetilde{H}]. \quad (12)$$

The boundary conditions to Eqs. (11) are

$$x'_1 - \left(\frac{b}{c}\right)x'_2 = q', \quad x''_1 + \left(\frac{b}{c}\right)x''_2 = q'',$$
$$p'_1 + \left(\frac{c}{b}\right)p'_2 = p', \quad p''_1 - \left(\frac{c}{b}\right)p''_2 = p''. \quad (13)$$

In order to integrate Eqs. (11) by Runge-Kutta methods for instance, we need the initial conditions $x_1(0)$, $p_1(0)$, $x_2(0)$ and $p_2(0)$. These are however unknown since we only have the propagator labels $q'$, $p'$, $q''$, $p''$ and the total time $T$. There is however a way. First we choose the initial variables $x_1(0)$ and $p_1(0)$ as the initial guess for each trajectory to be integrated. According to (13) we have

$$x_2(0) = \frac{c}{b}(x_1(0) - q'),$$
$$p_2(0) = \frac{b}{c}(p' - p_1(0)), \quad (14)$$

that is, given only the pair $(x'_1, p'_1)$, the complete set of initial conditions is determined. Integrating Eqs. (11) using these initial conditions led to a trajectory that most often fail to satisfy the *final* boundary conditions

$$x_1(T) + \left(\frac{b}{c}\right)x_2(T) = q'', \qquad (15)$$
$$p_1(T) - \left(\frac{c}{b}\right)p_2(T) = p''.$$

Let us define the function
$$D(x'_1, p'_1, T) =$$
$$\sqrt{\left[x_1(T) + \frac{b}{c}x_2(T) - q''\right]^2 + \left[p_1(T) - \frac{c}{b}p_2(T) - p''\right]^2} \qquad (16)$$

which is calculated after integrating Eqs. (11) under the initial conditions (14) and (15). This function actually measures the distance of the initial guess $(x'_1, p'_1)$ to the answer. Therefore, looking for an answer to the complex root problem under (13) was converted to the search of zeros of (16). The process is illustrated in Fig.1.

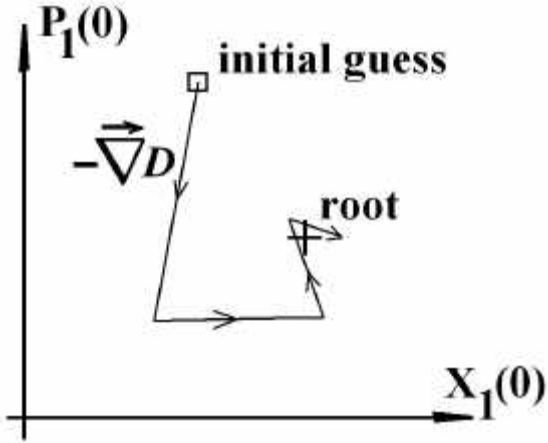

Fig. 1 Schematic representation of the search for zeros of $D$ in the space of initial guesses $(x'_1, p'_1)$

In the space of initial variables $(x'_1, p'_1)$, we start from an initial guess and calculate the gradient vector of $D$

$$\vec{\nabla} D = \left(\frac{\partial D}{\partial x'_1}, \frac{\partial D}{\partial p'_1}\right), \qquad (17)$$

which points to the negative of the decreasing values of $D$. The initial variables are updated at every j-th iteration step according to

$$x_1^{(j)}(0) = x_1^{(j-1)} - \mathbf{e}\frac{\partial D^{(j)}}{\partial x'_1},$$
$$p_1^{(j)}(0) = p_1^{(j-1)} - \mathbf{e}\frac{\partial D^{(j)}}{\partial p'_1} \qquad (18)$$

with $\mathbf{e}$ a small distance in the space $(x'_1, p'_1)$. In order to fasten convergence, the value of the gradient of $D$ is updated only when its value in a given iteration is greater than its previous one. Also, to attain accuracy and convergence, the value of the distance $\mathbf{e}$ should be decreased proportionally to $D$. Iteration continues until the value of $D$ is smaller than a certain $\mathbf{d} > 0$.

Before presenting some examples of the method at work, we give the relations of the SCSP as function of the dynamical variables $(x_1, p_1)$ which labeled the complex root for a given boundary condition. The prefactor, Eq. (9) contains the second derivative of $S_k$ which is

$$\frac{\partial^2 S}{\partial u' \partial v''} = \hbar\left[\left(\frac{b}{c}\right)\frac{\partial p''_1}{\partial q'} - \left(\frac{c}{b}\right)\frac{\partial x''_1}{\partial p'} - i\left(\frac{\partial x''_1}{\partial q'} + \frac{\partial p''_1}{\partial p'}\right)\right] \qquad (19)$$

After determining a complex trajectory, the function (19) is found by calculating the sensibility of the final trajectory coordinates $x''_1$ and $p''_1$ under changes in $q'$ and $p'$. Writing the complex action, Eq. (8) in the form

$$S(v'', u', T) = I_s + f,$$

with
$$I_s = \int_0^T \left[\frac{i\hbar}{2}(v\dot{u} - u\dot{v}) - \tilde{H}\right]dt,$$
$$f = -\frac{i\hbar}{2}(v''u'' + v'u'),$$

then
$$I_s = \frac{1}{2}\int[p_2 dx_2 + p_1 dx_1 - (x_1 dp_1 + x_2 dp_2)] +$$
$$+ i\frac{1}{2}\int[x_2 dx_1 + p_1 dp_2 - (x_1 dx_2 + p_2 dp_1)] +$$
$$+ [\tilde{H}_1(x_1, p_1, x_2, p_2) + i\tilde{H}_2(x_1, p_1, x_2, p_2)]T, \qquad (20)$$

since $\tilde{H}$ is a constant of motion for the trajectory and

$$f = \frac{1}{2}[q''p_1'' + p'x_1' - (p''x_1'' + q'p_1')] - \left[\left(\frac{c}{b}\right)(q''x_1'' + q'x_1') + \left(\frac{b}{c}\right)(p''p_1'' + p'p_1')\right] + \quad (21)$$
$$+ \frac{i\hbar}{2}(|z''|^2 + |z'|^2).$$

Finally, in Eq. (9), the phase of the exponential is

$$\int_0^T \frac{\partial^2 \tilde{H}}{\partial u \partial v} dt = \frac{1}{2}\int_0^T \left(b^2 \frac{\partial^2 \tilde{H}_1}{\partial x_1^2} + c^2 \frac{\partial^2 \tilde{H}_1}{\partial p_1^2}\right) dt - \frac{i}{2}\int_0^T \left(b^2 \frac{\partial^2 \tilde{H}_1}{\partial x_1 \partial p_2} + c^2 \frac{\partial^2 \tilde{H}_1}{\partial x_2 \partial p_1}\right) dt \quad (22)$$

where we use the fact that $\tilde{H}$ is an analytic function of $q$ and $p$.

## IV Examples of SCSP determination.

The complete determination of the SCSP in Eq. (5) also requires to find the phase $\mathbf{s}_k$ which arises from the phase of the amplitude term (19). A given complex function $z(t)$ can be represented in "catesian" notation

$$z(t) = a(t) + ib(t), \quad (23)$$

with $a(t)$ and $b(t)$ real functions. However, one can also write $z(t)$ in the form

$$z(t) = r(t)\exp(i\mathbf{a}(t) + in\mathbf{p}), \quad (24)$$

with $n = [0, 2, 4, \ldots]$ and the phase

$$\mathbf{a}(t) = \arctan\left(\frac{b(t)}{a(t)}\right). \quad (25)$$

Suppose that we take $\sqrt{z(t)}$. A phase factor arises in the polar representation since

$$\sqrt{z} = \sqrt{r}\exp\left(\frac{i\mathbf{a}(t)}{2} + \frac{in\mathbf{p}}{2}\right)$$

and the "true" phase is clearly underestimated since using (25)

$$-\frac{\mathbf{p}}{2} \leq \mathbf{a}(t) \leq \frac{\mathbf{p}}{2}.$$

So, in order to correctly retrieve the phase, we have to follow the sign of the functions $a(t)$ and $b(t)$. Let us call the "true" phase $\mathbf{j}(t)$ and further admit that $z(t)$ is a periodic function of $t$ with period $\mathbf{t}$. Table I gives the correct phase in terms of the sign of $a(t)$ and $b(t)$.

| $a(t)$ | $b(t)$ | $\mathbf{a}$ | $\mathbf{j}$ |
|---|---|---|---|
| $> 0$ | $\geq 0$ | $0 \leq \mathbf{a} < \mathbf{p}/2$ | $\mathbf{j} = \mathbf{a}$ (1) |
| $< 0$ | $\geq 0$ | $\mathbf{p}/2 < \mathbf{a} \leq \mathbf{p}$ | $\mathbf{j} = \mathbf{a} + \mathbf{p}$ (2) |
| $< 0$ | $\leq 0$ | $\mathbf{p} \leq \mathbf{a} < 3\mathbf{p}/2$ | $\mathbf{j} = \mathbf{a} + \mathbf{p}$ (3) |
| $> 0$ | $\leq 0$ | $3\mathbf{p}/2 < \mathbf{a} \leq 2\mathbf{p}$ | $\mathbf{j} = \mathbf{a} + 2\mathbf{p}$ (4) |

Table I Relation between the phase $\mathbf{a}$ and the true phase $\mathbf{j}$ of the complex number $z$ as function of the sign in $a(t)$ and $b(t)$. Four cases are possible.

Moreover, every time that $t > m\mathbf{t}$ with $m$ an integer number, the phase $\mathbf{j} \to \mathbf{j} + 2\mathbf{p}$. We take the number $z(t)$ as the complex function (19)

$$\frac{\partial^2 S}{\partial u' \partial v''} = a(T) + ib(T),$$

so that

$$\mathbf{s}(T) = \frac{1}{2}\mathbf{a}(T) + s\frac{\mathbf{p}}{2} + r\mathbf{p}, \quad (26)$$

with $s = 0, 1, 1,$ or $2$ according to the line case in Table 1 (1,2,3 or 4, respectively) and $r = 1, 2, 3, \ldots$ as $T$ is greater than a multiple of the period of function (19). Note that the phase (26) and the assumption of periodicity is valid for closed systems (such as quantum wells). In the case of open system (scattering problems) $r = 0$. Each complex trajectory contributing to (5) has its proper phase factor.

To illustrate the application of the method and the excellence of the SCSP in comparison to the exact calculation of the CSP we present some examples. We restrict to one dimensional Hamiltonians of the type (harmonic and quartic terms)

$$H(q, p) = \frac{1}{2}p^2 + \frac{1}{2}\mathbf{l}q^2 + \mathbf{b}q^4, \quad (28)$$

with $\mathbf{b} \geq 0$. If $\mathbf{b} = 0$ we obtain the simple harmonic oscillator for which the SCSP is exact. The smoothed Hamiltonian is easily found to be

$$\tilde{H} = \frac{1}{2}p^2 + \frac{1}{2}(\textbf{\textit{l}} + 6\textbf{\textit{bb}}^2)q^2 + \textbf{\textit{b}}q^4 + \frac{1}{4}(c^2 + \textbf{\textit{l}}b^2 + 3\textbf{\textit{bb}}^4), \quad (29)$$

where the "zero point" energy appears as a constant. The coherent state nature of the initial state also modifies the harmonic potential so that new harmonic frequency is

$$w = w_{classical}\sqrt{1 + \frac{\textbf{\textit{bb}}^2}{\textbf{\textit{l}}}}$$

in the presence of the quartic potential ($\textbf{\textit{b}} \neq 0$).

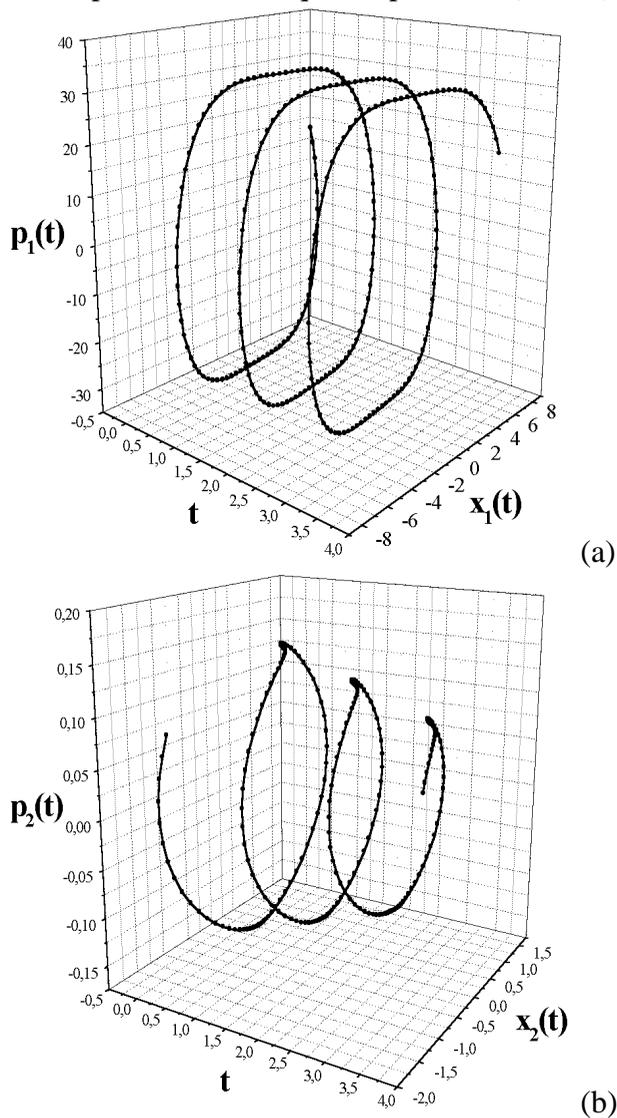

Fig.2 Perspective view in phase-time space of a complex trajectory in the quartic oscillator. (a) and (b) are the real and imaginary parts respectively.

In Fig.2 we see an example of complex trajectory calculated for the purely quartic system ($\textbf{\textit{l}} = 0$) and $\textbf{\textit{b}} = 0.2$ using the method exposed in section III. This figure shows the time evolution of the $(x_1, p_1)$ (part a) and $(x_2, p_2)$ (part b) along $t$. The CSP labels are $q' = 8.0$, $q'' = 6.0$, $p' = p'' = 15.0$ and $T = 3.5$. This orbit is close to a real trajectory for which $x_2(t) = p_2(t) = 0.0$ ($q' = q'' = 8.0, p' = p'' = 15.0$) in which case the primitive period is 1.003. Therefore in the $(x_1, p_1)$ graph, the trajectory shows three distinct turns along $t$ with a structure that closely resemble the phase-space trajectory of quartic systems. A similar time evolution is seen in the imaginary parts (b). The trajectory has 1000 points and the final value of the $D$ function was $8.6 \times 10^{-12}$. The initial coordinates which generate this trajectory are

$$x_1(0) = 6.68642954,$$
$$p_1(0) = 14.9244986.$$

In order to gain confidence in the accuracy of the SCSP, we must calculate the "exact" CSP and compare it to its semiclassical version. The exact CSP is easily obtained by its expansion in terms of the system eigenfunctions modulated by an amplitude factor that depends on the system eigen energies

$$K(z'', z', T) = \sum_n \langle z''|n\rangle\langle n|z'\rangle \exp\left(-\frac{iE_n T}{\hbar}\right). \quad (30)$$

In Eq.(30), $\langle z|n\rangle$ is the Husimi function of the n-th eingenstate and $E_n$ is the corresponding eigen energy. Both $|n\rangle$ and $E_n$ are easily calculated by diagonalizing the Hamiltonian (28) in a suitable basis [15] and performing the projection integrals into the coherent-state representation.

Using the relations presented in section III for the SCSP, we have calculated the time evolution of both conventional CSP and SCSP for different values of $\textbf{\textit{l}}$ and $\textbf{\textit{b}}$ using $\hbar = 1.0$ and $b = 1.0$. In Fig.3 we show the case $\textbf{\textit{l}} = 1.0$ and $\textbf{\textit{b}} = 0.01$. The figure shows the perfect agreement between the exact CSP (line) and its semiclassical counterpart (circles) for both real and imaginary components from $T = 0.0$ to $T = 10.0$. In this case $q' = q'' = 0.0$ and $p' = p'' = 1.0$. At every propagator time $T$, a single complex trajectory with 3000 points is used to calculate the SCSP until $D \leq 10^{-12}$. In order to obtain a faster convergence, the $(x'_1, p'_1)$ pair of a given trajectory is used as initial guess for the following one. In fact the number of time points between $T = 0.0$ and $T = 10.0$ is 1000 but we only show a sample with 50 in Fig. 3 and 4.

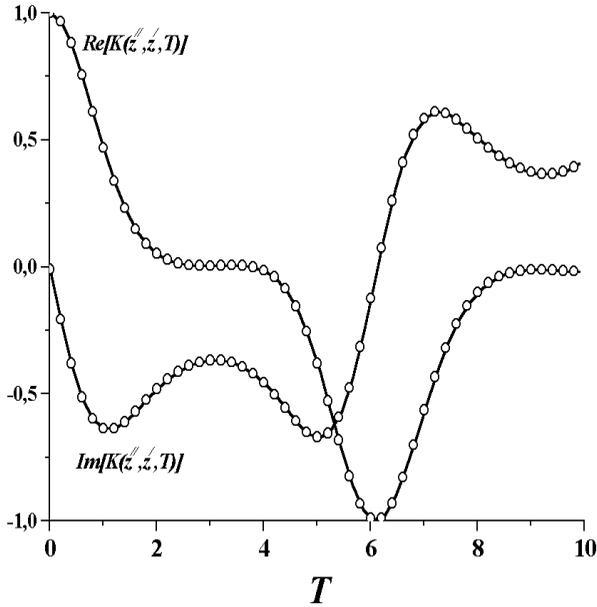

Fig. 3 Time evolution of the real and imaginary parts of the exact (lines) and SCSP (circles) for the harmonic + quartic oscillator. The agreement is excellent.

The agreement between the exact CSP and the SCSP for $\hbar = 1.0$ shows that the method is valid even outside the semiclassical range. As we said previously, the SCSP for the harmonic oscillator is exact, that is, it completely agrees with the CSP calculated by the conventional approach. This raises the question of whether the agreement in Fig. 3 could be explained by the perturbative (a harmonic oscillator slightly perturbed by a quartic term) character of the system at hand. To show that this is not the case, we calculated the time evolution of a purely non linear system with $l = 0.0$ and $b = 0.1$.

Fig. 4 shows the results from $T = 0.0$ to $T = 3.0$. Again the agreement between the real and imaginary parts of exact CSP and SCSP is excellent. This case was simulated with the same parameters of Fig. 3, $q' = q'' = 0.0$, $p' = p'' = 1.0$, $\hbar = 1.0$ and $b = 1.0$. The same convergence approach was used, each time point (along 1000 points) is built from *single* complex trajectory whose initial guess is taken from the previous $(x'_1, p'_1)$ pair.

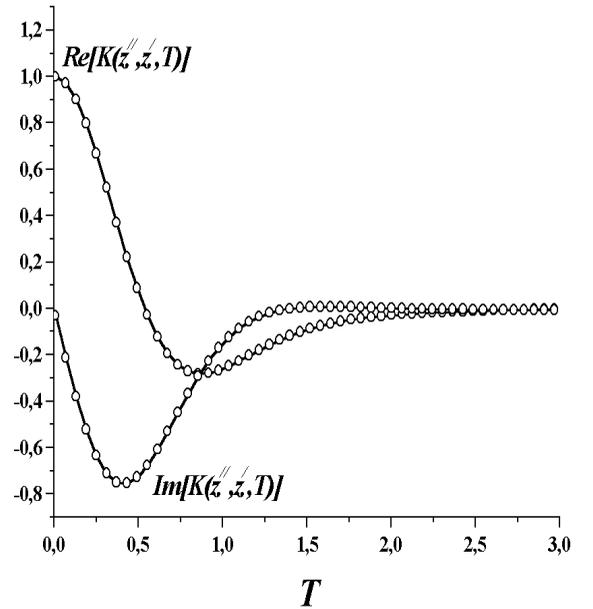

Fig. 4 Time evolution of the real and imaginary parts of the exact (lines) and SCSP (circles) for a purely quartic oscillator. The agreement is excellent

We would like to point out here that, in order to get the SCSP value for longer times in the quartic case, it is necessary to include contributions from several complex trajectories. These are generated by families of real trajectories at higher energies. Due to the dynamical character of the quartic potential, the shorter the period of its classical trajectories the higher their energies. Thus close to the propagator time $T = 0$, an infinite number of complex trajectories exist but their contributions are close to zero (the imaginary part of their complex action is very large) and the only relevant root is the so called "free" complex trajectory. Close to $T = 0$, the initial coherent packet behaves as a free particle and the SCSP is therefore constructed from correspondingly free complex paths.

## V Conclusion

We have shown here that the semiclassical approximation in the coherent-state representation leads to accurate values for the CSP amplitudes even out of the semiclassical range (large values of $\hbar$). The excellence and importance of the method is therefore confirmed. This importance is justified by the possibility of extending the semiclassical method to multi dimensional systems which would lead to interesting results mainly in the field of molecular dynamics.

The SCSP is based on a steepest descent procedure in which special roots are used to construct the semiclassical version of the coherent-state kernel. The roots are described as classical trajectories filling a complex phase space (both position and momentum are complex variables) that obey special boundary conditions. Usual classical dynamics is embedded in the complex one by the set of trajectories for which the imaginary position and momentum vanish. We have presented a new method based on a simple Runge-Kutta algorithm that allows a quick and easy calculation of complex trajectories. Special care should be taken to the phase factor [16] of the SCSP amplitude which arises from the amplitude factor of the semiclassical propagator.

The relative role of each complex trajectory in the SCSP amplitude (depending on the system) still need to be further clarified. In the case of the quartic potential, the number of contributing complex trajectories grows with $T$ as more and more turns of real trajectory families are involved. This is hardly the situation with other kinds of potentials whose orbital period dependence on energy is different . The role of complex trajectories is specially important in the semiclassical determination of spectra by the Fourier transform of CSP diagonal elements. A work in this direction will be presented elsewhere in the future.

## Acknowledgments

The author would like to acknowledge enlightening discussion with Marcus de Aguiar and the financial support by *FAPESP* under contract number 00/03168-0.